\documentclass[prl,a4paper,bm,twocolumn,noshowpacs,superscriptaddress]{revtex4}
\usepackage{bm}
\usepackage{amssymb}
\usepackage{amsfonts}
\usepackage{amsmath}
\usepackage{graphicx}
\usepackage[dvips]{color} 

\definecolor{g-blue}{rgb}{0.83,0.95,1}
\definecolor{Blue}{rgb}{0.5,0.5,1}
\definecolor{DarkBlue}{rgb}{0.00,0.00,0.58}
\definecolor{g-yellow}{rgb}{1,1,0.7}
\definecolor{g-green}{rgb}{0.9,1,0.9}
\definecolor{green}{rgb}{0,0.6,0}
\definecolor{Green}{rgb}{0,0.4,0}
\definecolor{cyan}{rgb}{0,0.7,0.7}
\definecolor{black}{rgb}{0,0,0}
\definecolor{grey}{rgb}{0.4 ,0.4 ,0.4 }

\def\Fbox#1{\vskip1ex\hbox to 8.5cm{\hfil\fboxsep0.3cm\fbox{%
  \parbox{8.0cm}{#1}}\hfil}\vskip1ex\noindent}  

\def\be{\begin{equation}}\def\ee{\end{equation}}
\def\bea{\begin{eqnarray}}\def\eea{\end{eqnarray}}
\def\bse{\begin{subequations}}\def\ese{\end{subequations}}
\newcommand{\BE}[1]{\begin{equation}\label{#1}}
\newcommand{\BEA}[1]{\begin{eqnarray}\label{#1}}
\newcommand{\BSE}[1]{\begin{subequations}\label{#1}}

\let \= \equiv \let\*\cdot \let\~\widetilde \let\^\widehat \let\-\overline

  \def\1{\bm1} 

\def\<{\left\langle}    \def\>{\right\rangle}
\def\({\left(}          \def\){\right)}
 \def \[ {\left [} \def \] {\right ]}

 \renewcommand{\O}{\Omega}

\newcommand{\C}[1]{{\mathcal{#1}}}    

\renewcommand{\sb}[1]{_{\text {#1}}}  

\newcommand{\vf}{V_{\mathrm f}}

\def\Oms{\Omega_{\rm s}}
\def\Vf{V_{\rm f}}
\def\aam{\alpha_{\rm am}}
\def\aen{\alpha_{\rm en}}
\def\taam{\tilde\alpha_{\rm am}}
\def\taen{\tilde\alpha_{\rm en}}
\def\Cam{C_{\rm am}}
\def\Cen{C_{\rm en}}

\begin{document}

\title{Energy and angular momentum balance in
wall-bounded quantum turbulence at very low temperatures}

\author{J.J.~Hosio$^*$} \author{V.B.~Eltsov}  \author{P.J. Heikkinen}  \author{R. H\"anninen} \author{M.~Krusius}

\affiliation{O.V. Lounasmaa Laboratory, Aalto University, P.O. Box 15100, FI-00076 AALTO, Finland}
\author{V.S.~L'vov}
\affiliation{Department of Chemical Physics, The Weizmann Institute of
  Science, Rehovot 76100, Israel}



\begin{abstract}
A superfluid in the absence of the viscous normal component should be the best realization of an ideal inviscid Euler fluid. As expressed by d'Alembert's famous paradox, an ideal fluid does not exert drag on bodies past which it flows, or in other words, it does not exchange momentum with them. Also, the flow of an ideal fluid does not dissipate kinetic energy. We study experimentally whether these properties apply to the flow of superfluid $^3$He-B in a rotating cylinder at low temperatures. It is found that ideal behavior is broken by quantum turbulence, which leads to substantial energy dissipation, as observed also earlier. Here we show that remarkably, nearly ideal behavior is preserved with respect to the angular-momentum exchange between the superfluid and its container, i.e., the drag almost disappears in the zero-temperature limit. This mismatch between energy and angular-momentum transfer results in a new physical situation where the proper description of wall-bounded quantum turbulence requires two effective friction parameters, one for energy dissipation and another for momentum coupling, which become substantially different at very low temperatures.
\end{abstract}
\maketitle 

A remarkable property of incompressible and inviscid potential flow, as discovered by d'Alembert in mid 18th century, is that bodies moving at constant velocity relative to this ideal fluid experience no drag \cite{landau_fluid}. This gross contradiction with observations of real fluids was resolved in the beginning of the 20th century by Prandtl, who introduced the concept of a boundary layer obeying  the no-slip boundary condition. Rapid spatial variation of the velocity in the boundary layer sustains viscous effects which lead to drag even in flow at high Reynolds number.

In superfluids at temperatures much below the critical temperature $T_{\rm c}$, the viscous normal component is almost absent. The superfluid component might seem to be close to an ideal fluid considered by d'Alembert, both in the bulk, where vortex-free flow is potential, and at the boundaries, where the no-slip condition of viscous fluids reduces to the 'no-flow-through-the-boundary' condition. In reality, however, the dynamics of different non-trivial superfluid flows includes also the motion of quantized vortex lines. In many cases, this takes the form of quantum turbulence, i.e., the complex motion of tangled and reconnecting vortices. It has been demonstrated recently that quantum turbulence breaks the correspondence between superfluids and ideal fluids. In particular, finite energy dissipation, and thus finite effective viscosity or friction, has been observed in low-temperature helium superfluids \cite{mcclintock,PhysRevLett.96.035301,front1,Manchester} and was in some cases linked to a quasiclassical turbulent energy cascade \cite{Manchester,lanc_nat}.

\begin{figure}
\centerline{\includegraphics[width=\linewidth]{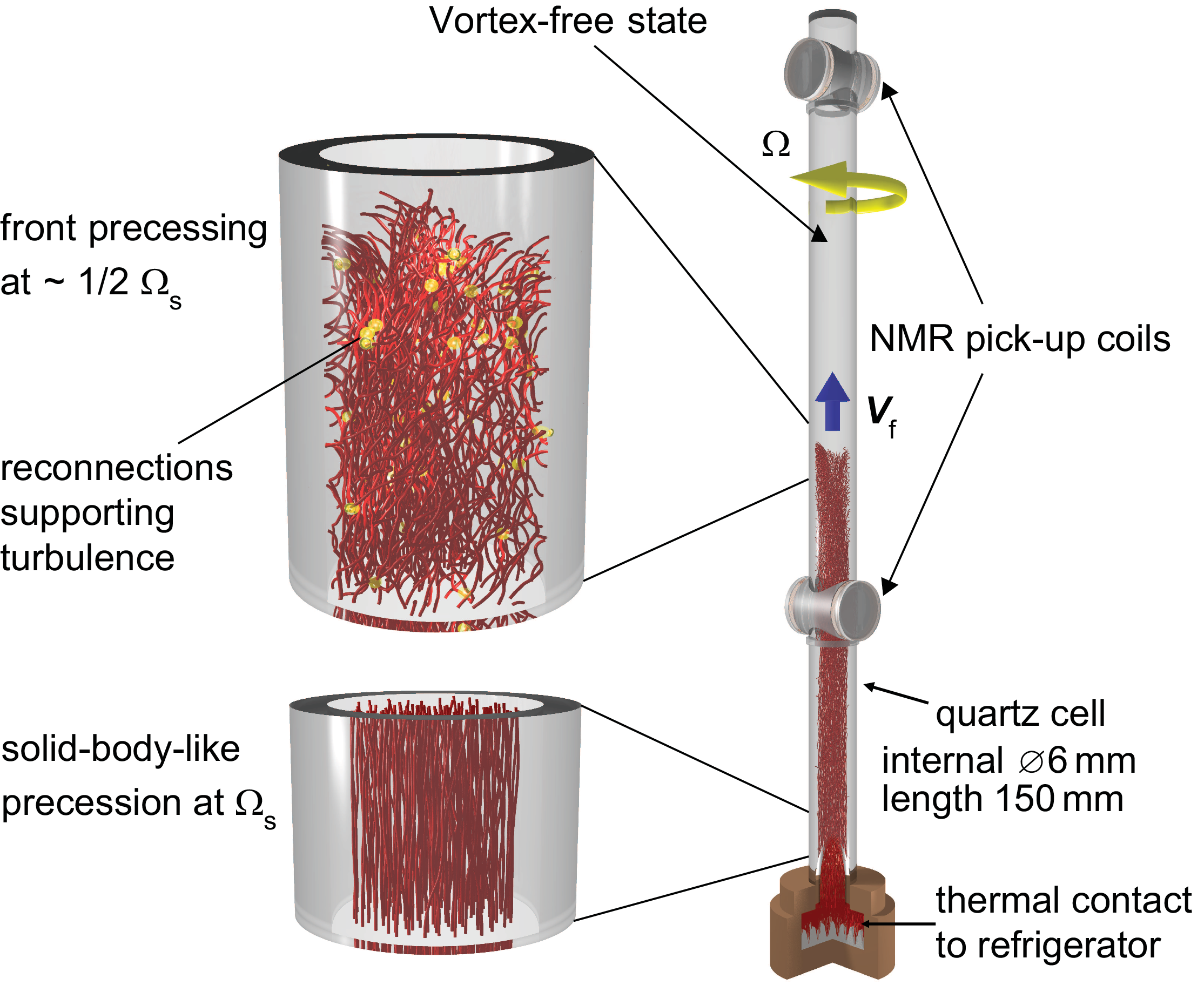}}
\caption{\textbf{Turbulent vortex front motion in a rotating cylinder.} The front moves axially upward and rotates azimuthally with respect to the cylinder. The motion is detected with two NMR pick-up coils, which are 9~cm apart. In the front, where the vortices bend perpendicular to the side wall, the angular velocity of the superfluid and also the precession frequency of the vortices changes from zero to $\Oms \lesssim \Omega$. The difference in the average precession frequencies between the front and the vortex bundle behind it enforces reconnections (yellow dots) and turbulence in the front, as seen in the zoomed view on the top left. The differential precessions also wind the vortex lines behind the front to a twisted configuration \cite{twist_prl}. This snapshot of the ensuing vortex configuration comes from numerical calculations at
  $0.27\,T_\mathrm{c}$. }
\label{setup}
\vspace{-3mm}
\end{figure}

\begin{figure*}[t!]
\centerline{\includegraphics[width=0.9\linewidth]{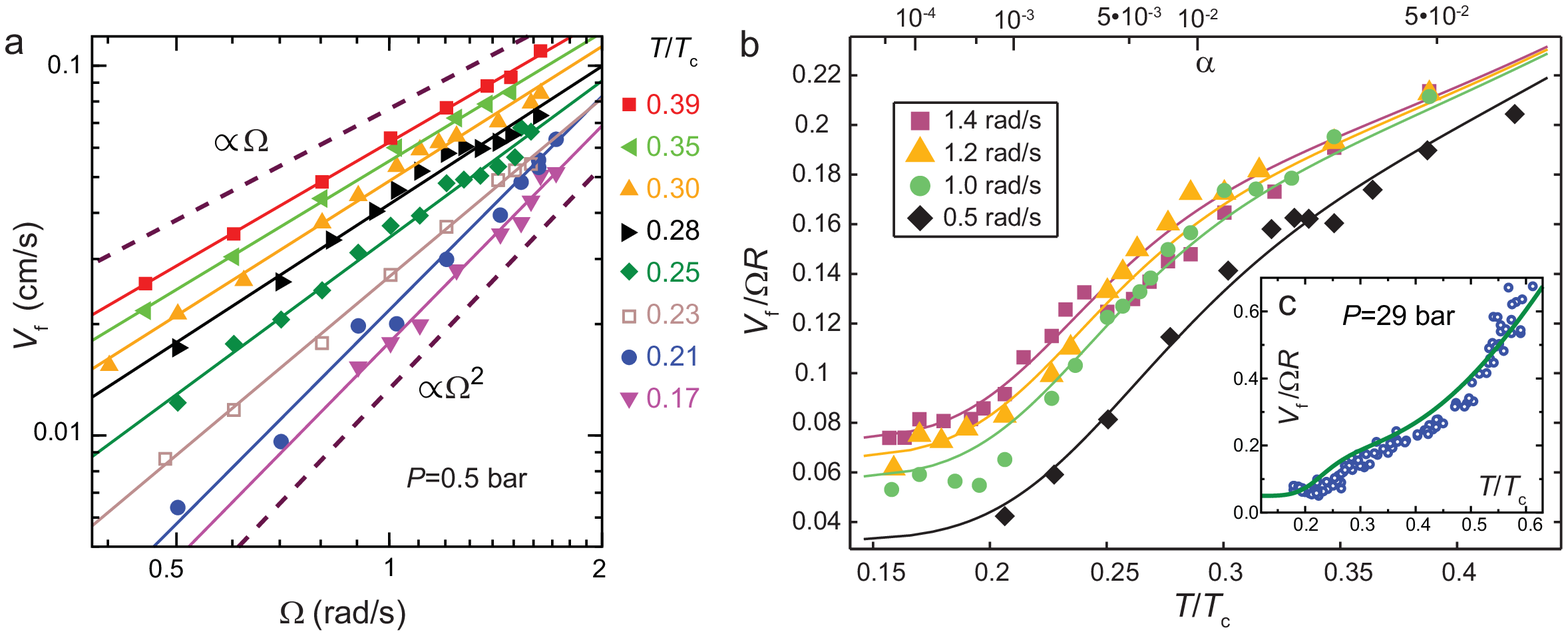}}
\caption{\textbf{Axial front velocity and fits to the model Eq.~(\ref{model})}. \textbf{a,} Velocity $\vf(T,\O)$ plotted vs. the angular velocity of rotation $\Omega$, with logarithmic axes. The dashed lines correspond to $V_\mathrm{f}\propto\O$ and $V_\mathrm{f}\propto\O^2$ for comparison. \textbf{b,} Dependence of $V_{\rm f}/(\O R)$ on temperature $T$ at four different $\Omega$.  In the $T\to 0$ limit $V_{\rm f}$ tends to a $T$-independent but $\Omega$-dependent value determined by the residual terms $\taen$ and $\taam$. On the top horizontal axis the mutual friction $\alpha (T)$ is given \cite{prec}. \textbf{c,} In the inset the data on $\vf(T)$ measured at $29$\,bar pressure in Ref.~\cite{front1} for $\O \approx 1$~rad/s is compared to Eq.~(\ref{model}), using the measured $\alpha (T)$ for 29\,bar \cite{PRL10} and the same values for the four fitting parameters as obtained from the present 0.5~bar measurements.
}
\label{sfv}
\vspace{-3mm}
\end{figure*}

However, an intriguing question is whether low-temperature quantum turbulence also mimics viscous boundary flow by producing similar drag. So far measurements of the superfluid drag in the $T\rightarrow 0$ limit have been performed only for flows around various oscillating objects \cite{schoepe,yano,PhysRevB.85.224533}. In these experiments, the ideal boundary conditions have not been necessarily satisfied, owing to surface pinning, and substantial drag is measured, approaching in magnitude that of the normal fluid \cite{PhysRevB.85.224533}. Our measurement of energy dissipation and momentum exchange with the container walls in steady-state turbulent flow is sketched in Fig.~\ref{setup}. We measure the motion of a turbulent vortex front propagating along a rotating cylinder filled with superfluid $^3$He in the B phase. We use a container with very smooth walls, which means that practically pinning is excluded and ideal flow boundary conditions are restored. This is possible in $^3$He-B owing to the large vortex core radius ($a\sim 0.1\,\mu$m at 0.5\,bar liquid pressure).

As a result we find that a boundary layer similar to that in classical
fluids is not created. In the front motion, exchange of momentum with the
walls turns out to be significantly suppressed in comparison to energy
dissipation. This discrepancy between nearly-ideal-liquid behaviour with
respect to drag and dissipative behaviour with respect to energy transfer has
profound implications on the $T \rightarrow 0$ dynamics of quantum fluids.
In particular, superfluid decouples from the motion of the container to
reduce the amount of the transferred momentum. Additionally, the role of
the slow laminar dynamics, which allows for momentum exchange with the walls
via the diluted normal component, substantially increases compared to what one
would expect from the higher-temperature behaviour.

\section{Results}
\textbf{Axial velocity of vortex front.} The measurement starts with the container at rest, with vortex-free non-rotating superfluid. When the container is set into rotation at an angular velocity $\Omega$, a turbulent front forms at the rough surfaces of the heat exchanger at the bottom and starts to move upwards along the cylinder, bringing the superfluid behind it into rotation \cite{front1}. The axial front velocity $\Vf$ is determined from the flight time between two pick-up coils of nuclear-magnetic-resonance (NMR) spectrometers. The energy dissipation rate can be inferred from $\Vf$ by considering the free energy difference of the superfluid before and after the front. The validity of this approach was proven by direct calorimetric measurement of the dissipated heat in the front propagation \cite{front2}. The rate of the angular-momentum exchange with the bounding walls can be determined either directly from the rotation $\Oms$ of the superfluid behind the front or indirectly from the dependence of $\Vf$ on $\Omega$. We first discuss the indirect approach.

The measured axial front velocity is presented in Fig.~\ref{sfv} as a function of $\Omega$ at constant temperature and as a function of temperature at constant $\Omega$. As seen from Fig.~\ref{sfv}a, the $\Omega$-dependence of $V_\mathrm{f}$ gradually changes from linear $\Vf \propto \O$ towards quadratic $\Vf \propto \O^2$ with decreasing temperature below 0.3\,$T_{\rm c}$. From Fig.~\ref{sfv}b one finds the limiting behaviour as a function of temperature: at high temperatures the normalized front velocity approaches the $\O$-independent 'single-vortex result' $\vf /(\O R) \approx \alpha (T)$ \cite{front1} and below $0.2\,T_{\mathrm{c}}$ it appears to level off to a $T$-independent but $\O$-dependent value.  Here $R$ is the radius of the cylinder and $\alpha(T)$ is the dissipative mutual friction parameter, which characterizes the coupling of the normal and superfluid components of $^3$He-B via scattering of the thermal quasiparticle excitations from the vortex cores \cite{kopnin}. At the lowest temperatures, the inter-quasiparticle collisions are absent, the excitations move along ballistic flight paths, and the friction is exponentially reduced as $\alpha\propto \exp(-\Delta/T)$, where $\Delta$ is the superfluid energy gap.

\textbf{Phenomenological model of front propagation.} The front propagation can be understood using a phenomenological model based on the consideration of energy and angular-momentum balance in the front. The model relies on the equation of superfluid hydrodynamics averaged over vortex lines \cite{sonin_RMP}, which is basically the Euler equation with an additional friction force from the normal component and a line-tension force owing to the quantum nature of vortices. The friction force is parametrised with $\alpha(T)$, while the tension-force parameter is $\lambda = (\kappa/4\pi)\ln(\ell/a)$, where $\kappa = h/(2 m_3) \approx 0.07\,$mm/s$^2$ is the quantum of circulation and $\ell \approx (2 \Oms /\kappa)^{-1/2} \approx 0.1\,$mm is the inter-vortex distance. Strictly speaking, these equations are applicable only when the vortex lines are locally roughly parallel to each other. To allow an extension to the case of the turbulent front, we replace the friction $\alpha$ with two separate effective friction parameters: $\aen$ when we consider the energy balance and $\aam$ for the angular momentum balance. These parameters are defined as follows:
\begin{equation}
\aen = \Cen \alpha(T) + \taen,\quad
\aam = \Cam \alpha(T) + \taam.
\label{aleff}
\end{equation}
Here $\Cen$, $\Cam \sim 1$ account for the modification of the friction from the normal component owing to the reduced polarization of the vortex lines and their fluctuating motion in the turbulent front. The parameters $\taen$ and
$\taam$ describe the contribution of turbulence and possible effects from residual surface friction. For simplicity, we consider all four phenomenological parameters to be $(T,\Omega)$-independent.

Analysis of the energy and the angular-momentum balance (see the Supplementary Discussion for details) leads to the following expressions for the front velocity $\Vf$ and the angular velocity $\Oms$  of the superfluid behind the front:
\begin{equation}
\Vf = \aen \Oms R, \qquad
\Oms= \frac{\alpha_{\rm{am}} \Omega^2}  {\alpha_{\rm{am}}\Omega+\lambda R^{-2}}.
\label{model}
\end{equation}
The expression for $\Vf$ is a direct generalization of the laminar single-vortex result. The expression for $\Oms$ follows from balancing the azimuthal components of the friction and line-tension forces. The friction force appears since the azimuthal velocities of the normal and superfluid components differ, while the line-tension force is due to the twisting of the vortices behind the front.

The expression for the front velocity $\Vf$ (Eq.~\ref{model}) is compared in Figs.~\ref{sfv}a,b to its measured values in the temperature range, where the mutual friction $\alpha(T)$ changes by roughly three orders of magnitude. The solid lines show the model (\ref{model}) with the four fitting parameters $\Cen=0.52$, $\Cam=1.33$, $\taen=0.20$, and $\taam=0.0019$. The
temperature dependence $\alpha(T)$ is taken from the measurements in Ref.~\cite{prec}. The model seems to be in good agreement with the measurements and in particular, reproduces all the qualitative features of $\Vf(T, \Omega)$ discussed above.

\begin{figure}[t!]
\includegraphics[width=0.9\linewidth]{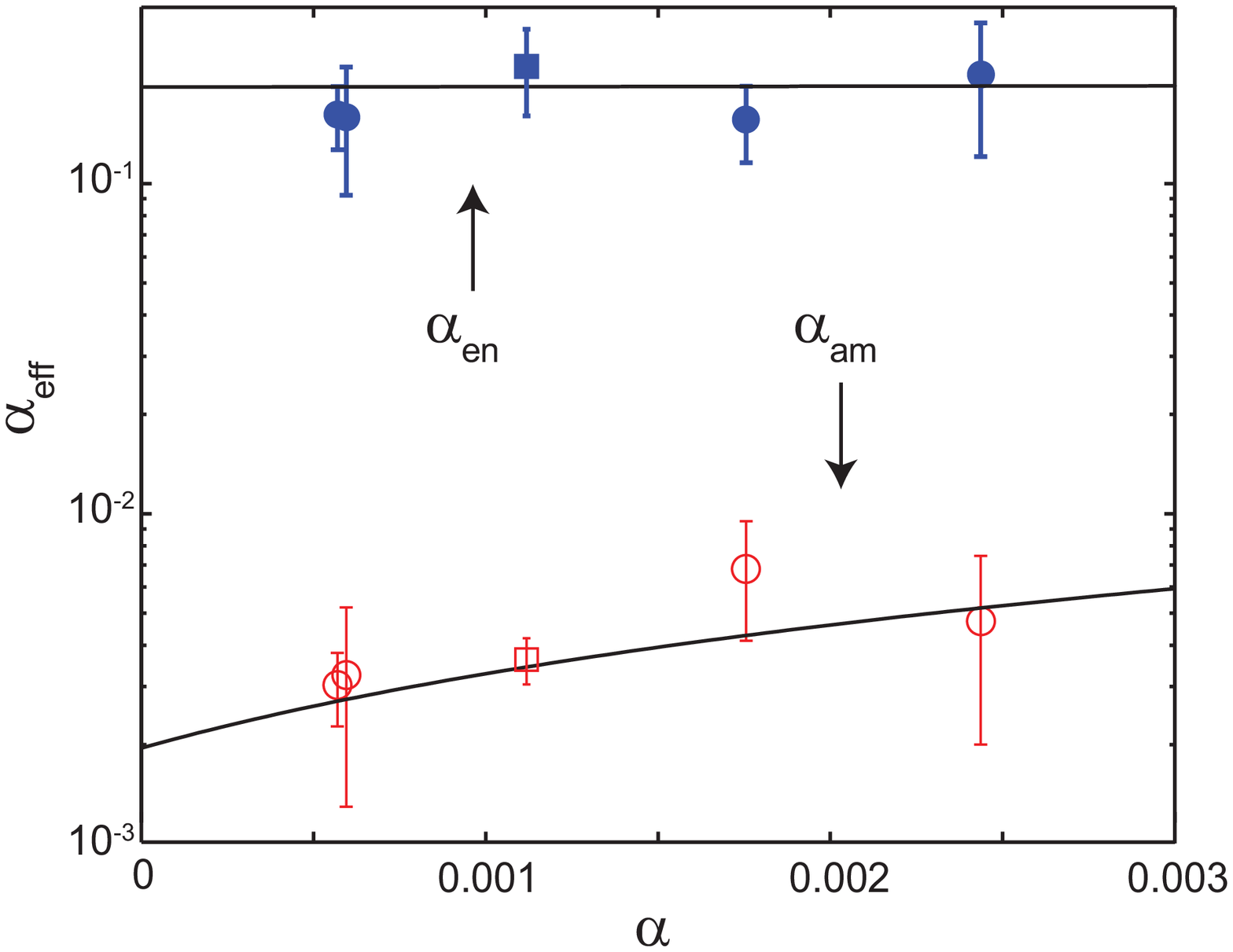}
\caption{\textbf{Direct measurement of the effective friction parameters and comparison to the model.} The data points $\aam$ and $\aen$ have been extracted from direct simultaneous measurements of azimuthal precession $\Oms (T, \O, P)$ and axial front propagation velocity $\vf (T, \O, P)$ using Eq.~(\ref{model}) and are plotted as function of temperature in terms of the measured mutual friction $\alpha (T,P)$. The superfluid angular velocity $\Oms$ is measured from the precession frequency of the vortex bundle behind the front, with $\O$ in the range 0.4 -- 1.0\,rad/s using the method described in Ref. \cite{front2}. Circles denote measurements at $P = 0.5\,$bar and squares at 29\,bar. The solid lines represent Eq.~(\ref{aleff}) with the same parameters as used to fit the velocity data in Fig.~\ref{sfv}.}
\label{aeff}
\end{figure}

\textbf{Direct measurement of superfluid angular velocity.} Occasional small axial asymmetries in the precessing front and in the bundle behind it cause oscillations in the NMR signal \cite{front2}. In such cases, the superfluid angular velocity $\Oms$ can be directly extracted from the frequency of the oscillations, albeit usually in a limited range of temperatures slightly above 0.2\,$T_{\rm c}$. This allows us to determine the angular-momentum coupling parameter
$\aam=\lambda\Oms/(R^2\Omega(\Omega-\Oms))$ independently of the front velocity $\Vf$. The results shown in Fig.~\ref{aeff} highlight again the good agreement with the fitted parameter values extracted from Fig.~\ref{sfv} and the two orders of magnitude difference between $\aen=\vf/\Oms R$ and $\aam$.

This difference and in particular the minute value of $\taam$ emphasize our central result. When the influence of the normal component on the energy dissipation and coupling to the container walls is excluded, $\alpha \to 0$, the residual effective friction parameters $\taen$ and $\taam$ should be primarily associated with quantum turbulence in the front. The relatively large value of $\taen \approx 0.2$ shows that turbulence in the front efficiently enhances energy dissipation. For comparison, the same friction from the normal component, $\alpha \approx 0.2$, is reached at $T\gtrsim0.45 T_{\rm c}$ where the density of the normal component is about 30\% of the total liquid density. The contribution of quantum turbulence to the dissipation in the zero-temperature limit is usually attributed to the energy transfer in the turbulent cascade to length scales much smaller than the inter-vortex distance, where it is dissipated by the emission of non-thermal quasiparticles from the vortex cores \cite{silaev}.

The much smaller residual term $\taam \ll \taen$ attests that the angular-momentum transfer from the boundaries is much less effective and that the drag from the superfluid on the boundaries almost disappears in the zero-temperature limit.
It is not clear whether the value of $\taam$ should be completely ascribed to the residual friction in the motion of the surface-attached vortices or whether it has a contribution from turbulence such as momentum transfer from cascading effects similar to those in the energy dissipation or via annihilation of small vortex loops at the surfaces.
There are some arguments, though, in favor of the turbulent contribution.
First, the value of $\taam$ is four times larger than the residual friction term in laminar vortex motion \cite{prec}. Also Figs.~\ref{sfv}c and \ref{aeff} demonstrate that the same fitted effective friction parameters explain earlier high-pressure measurements \cite{front1} which were conducted in a different quartz cylinder. We take this to indicate that the model parameters are not strongly dependent on our surfaces or on the pressure-dependent vortex core radius.

\section{Discussion}
To summarize, we find that two separate effective friction parameters have to be introduced to account for energy and angular-momentum transfer respectively. By comparing the measurements to a phenomenological model of turbulent front propagation, it is seen that these parameters differ by two orders of magnitude. This difference leads to new physical effects in superfluid dynamics in the $T\rightarrow0$ limit, such as the recently observed decoupling from the reference frame of the container \cite{front2}. Decoupling of the superfluid and normal components is also discussed, for instance in the evolution of a rotating neutron star when it cools through $T_\mathrm{c}$ \cite{pulsars}, but here in the $T\rightarrow0$ limit the mechanism of decoupling is different. It leads to the unusual situation that efficient energy dissipation is achieved in turbulent bulk flow, but the angular-momentum exchange is not enhanced as much beyond that provided by the bulk mutual friction coupling and thus a turbulent boundary layer, similar to that in the classical case, is not formed.

\vspace{3mm}

\section{Methods}\small{
\textbf{Sample preparation and triggering vortex motion.} The measuring setup consists of a long cylinder made from fused quartz and filled with liquid $^3$He-B at 0.5\,bar pressure. The internal surfaces of the cylinder are treated with hydrofluoric acid \cite{HFetching} and are appropriately cleaned to remove pinning sites for vortices. At its bottom end, the cylinder opens to a sintered-silver heat exchanger. Its rough surface ensures that vortices are formed there at low rotation velocity. In the experiment, we rapidly increase the rotation velocity from zero to the desired target velocity $\Omega$. This creates a turbulent burst at the bottom end and triggers an upward-propagating vortex front. Owing to the smooth walls, the critical velocity of vortex formation is as high as 1.8\,rad/s elsewhere in the quartz cylinder and thus at lower velocities, the volume above the front remains in the meta-stable vortex-free Landau state, which eventually is displaced by the front. The final state, which is reached only long after the front has reached the end of the cylinder via laminar spin-up of vortices \cite{front3}, is the equilibrium vortex state with a constant solid-body density of rectilinear line vortices. After each measurement, rotation is stopped and the sample is warmed up to $\sim0.7\,T_{\rm c}$ for about one hour, to allow remanent vortices to annihilate \cite{vort_ann}, before cooling down again for a new measurement. Alternatively, the sample may be warmed up above $T_{\rm c}$ upon which it can be cooled down immediately.

\textbf{NMR detection techniques and thermometry.} The propagation of the front is monitored with two NMR pick-up coils. At temperatures above 0.2\,$T_{\rm c}$ and velocities above 0.6\,rad/s, the arrival of the front can be detected by tracing the NMR signal at the so-called counterflow peak in the linear NMR spectrum \cite{deGraaf}. The arrival is seen as a rapid decrease of the NMR absorption. This method loses sensitivity at lower rotation velocities and temperatures, as the counterflow peak rapidly decreases in amplitude. There the front propagation is monitored from the frequency of the coherent non-linear NMR signal generated by a Bose-Einstein condensate of magnon quasiparticles in a magnetic trap in the middle of the pick-up coil. This frequency depends strongly on the difference in the velocities of the azimuthal flow of the normal and the superfluid components which provides a contribution to
the radial trapping of the magnon condensate \cite{magnon}. A small axial pinch coil is used around the NMR pick-up coil to provide the localization of the trap in the axial direction.  The temperature is determined from the resonance width of a quartz tuning fork oscillator, located close to the heat exchanger, which depends on temperature as $ \Delta f=11460
{\rm{Hz}} \exp (-1.776T_{\rm c}/T)$ in the ballistic regime of quasiparticle transport.}


\section{Acknowledgements}
We thank E. Kozik, E.B. Sonin, and G.E. Volovik for stimulating discussions.
The work is supported by the Academy of Finland (Centers of Excellence Programme 2012-2017 and grant 218211), the EU 7th Framework Programme (FP7/2007-2013, grant 228464 Microkelvin), and the USA-Israel Binational Science Foundation. J.J.H. and P.J.H. acknowledge financial support from the V\"{a}is\"{a}l\"{a} Foundation of the Finnish Academy of Science and Letters.

\section{Contributions}
The experiments were carried out by J.J.H. and P.J.H. and the data were analyzed by  J.J.H. and V.B.E. V.S.L. supplied theoretical background and set up the model for the explanation of the experiments. R.H. provided the illustrations on vortex structures based on his numerical simulations. The manuscript was written by J.J.H., V.B.E., V.S.L., and M.K. with contributions from all authors. J.J.H., V.B.E., and M.K. did the devising of the experiment. All the authors contributed to the discussion of the results. V.B.E. supervised the project.

{\noindent \section{Supplementary Discussion}}

Large scale superfluid turbulence can be analyzed  in the framework of the coarse-grained hydrodynamical equation for the superfluid velocity~[12,23]:
\begin{subequations}
\label{Hydrodynamics}
\begin{eqnarray*}\label{EE} && {\partial {\bf v}_{\rm s} }/{ \partial t}+
  \nabla(\mu + v_{\rm s}^2/2) = {\bf v}_{\rm s} \times (\nabla\times  {\bf v}_{\rm s})   + {\bf F}_{\alpha} + {\bf F}_{\lambda}\,, \\ \label{mf} &&  {\bf
  F}_{\alpha}= - \alpha(T)~\hat{\bf \omega} \times(( {\bf v}_{\rm s}- {\bf v}_{\rm n})  \times (\nabla\times {\bf v}_{\rm s}) )\,,\\ \label{lt} && {\bf
  F}_{\lambda}= - \lambda (\nabla\times {\bf v}_{\rm s})\times (\nabla\times \hat{\bf \omega})\ . \end{eqnarray*}
\end{subequations}
Here  $\mu$ is the chemical potential,  ${\bf F}_{\alpha}$ is the dissipative mutual-friction force, and  ${\bf F}_{\lambda} $ is the vortex-line-tension force.
The unit vector $\hat{\bf \omega}$ is directed along the vorticity $\nabla\times {\bf v}_{\rm s}$. The line-tension parameter, $ \lambda\equiv
(\kappa/4\pi)\ln(\ell/a)$, depends on the intervortex distance, $\ell \approx \sqrt{\kappa/2\Omega}\sim 0.1\,$mm (at $\Omega\sim 1\,$rad/s), and on the vortex
core diameter, $a \approx 70$~nm (at 0.5 bar liquid pressure).

\textbf{Global balance of energy. } The total rate of energy dissipation $d \C E_-/ dt $ (per unit mass) in stationary turbulent front propagation is equal to the
total free energy input [17]
\begin{equation*}
\label{der1b} d {\cal E}_+/ dt = -\pi   V\sb f R^4\Omega\sb s \big (2\, \Omega -\Omega \sb s \big)/4\ .
\end{equation*}
This expression is the free energy  difference  before the front (where ${\bf v}_{\rm s} = 0$) and after the front (where ${\bf v}_{\rm s} \approx
\mathbf{\Omega\sb s} \times \mathbf{r}$) times the rate of volume variation. For simplicity, we present here a dimensional reasoning omitting numerical
prefactors. Dimensionally $d {\cal E}_- / dt $ can be written as follows:
\begin{equation*}
\label{der1c} d {\cal E}_-/ dt = - \pi  R^5\Omega_1 \Omega_2 \Omega_3 F (\Omega\sb s/ \Omega)\,,
\end{equation*}
where $\Omega_1$,  $\Omega_2$, $\Omega_3$ are some frequencies and $F(\Omega\sb s/ \Omega)$ is dimensionless function of dimensionless argument. Equating  $d
{\cal E}_+/ dt$ and $d {\cal E}_-/ dt$ one gets the  equation for the front velocity:
\begin{equation*}\
\label{der1d} \displaystyle V\sb f= \frac{\Omega_1 \Omega_2 \Omega_3\, R }{\Omega\sb s \big (2\, \Omega -\Omega \sb s \big)}  \,   F \Big(\frac{\Omega\sb s}{
\Omega}\Big )  \ .
\end{equation*}

It is intuitively clear that for $\Omega\sb s\ll \Omega$ the front velocity should be of order $\sim \Omega\sb s  R$ and go to zero in the limit $\Omega\sb s\to
0$. This can be reached by putting $\Omega_1=\Omega_2=\Omega_s$. Another restriction follows from the fact that for $\Omega\sb s= 2 \Omega$  there is no reason
for the front velocity to diverge. This can be prevented by putting $\Omega_3=2\, \Omega-\Omega\sb s$. If so, the front velocity becomes:  \begin{equation*}
\label{der1e} V\sb f=\Omega\sb s R\,   F ( \Omega\sb s/\Omega   )\,, \end{equation*} with function $F(y)$ finite for $y=0$ and $y=2$.

At temperatures above $0.4\,T_{\rm c}$, front motion is laminar and its velocity can be found analytically ~\cite{finne,sonin_front}. The resulting equation $
V\sb f=\alpha (T)\Omega\sb s R$ dictates that in the  laminar front $F(y)= \alpha (T)$. At lower temperatures, where the dissipative mutual-friction parameter
$\alpha(T)$ decreases exponentially with temperature, front motion starts to deviate from laminar behavior owing to the extra dissipation from turbulence [4].
This effect can be accounted for by replacing in the energy balance [with $F(y)=\alpha(T)$]  the laminar friction parameter $\alpha(T)$ with an effective
turbulent mutual friction

\begin{equation*} \vf \approx \alpha_{\rm{en}} (T)\,  \Omega\sb s R\,, \quad \alpha_{\rm{en}}(T)=\Cen \alpha (T)+\taen \, .~~~ \label{en} \tag{S1}
\end{equation*}
The first term in $\alpha_{\rm{en}}(T)$ originates from mutual-friction dissipation, where $\Cen$ is of order unity. The second term $\taen$ describes residual
turbulent energy dissipation in the energy cascades toward small scales. In principle, at the scale of the inter-vortex distance it depends on $T$ and $\Omega$,
as discussed for the bottleneck in the turbulent energy cascades in Refs.~[4,26,27].

\textbf{Global balance of angular momentum}. Since the transfer of angular momentum within the turbulent front is more effective than the coupling of the
superfluid to the container walls, we consider only the global force balance for the azimuthal components $\langle F_{\alpha} \rangle_\phi = \langle F_{\lambda}
\rangle_\phi $: the former controls the spin-up due to the coupling to the normal component, while the latter describes the tendency of the superfluid to decrease
the vortex number due to the vortex-line tension between the vortices in the front and the bundle behind it. With the simple estimate $\nabla\times {\bf v}_{\rm
s} \simeq 2 \Oms$ one obtains
$$\langle F_{\alpha} \rangle \simeq 2\alpha \Oms(\Omega-\Oms)R.$$
The azimuthal line-tension force vanishes in the solid-body approximation (when ${\bf v}_{\rm s} =  \mathbf{\Omega\sb s} \times \mathbf{r}$). However, since the
vortices in the bundle are nonuniformly twisted, the vorticity $\hat{\mathbf{\omega}}$ is not perfectly parallel with $\hat z$. The twist is proportional to
$\Oms$ and has to be normalized by $\Omega R$, i.e., $\nabla\times {\bf{\hat\omega}} \simeq \Oms/(\Omega R)$ yielding
$$\langle F_{\lambda} \rangle \simeq 2\lambda \Oms^2/(\Omega R).$$
Now the force balance gives
\begin{equation*} \alpha_{\rm{am}} \Oms(\Omega-\Oms)R=\frac{\lambda \Oms^2}{\Omega R}\,, \quad  \alpha_{\rm{am}}=C_{\rm{am}}\alpha(T)+\taam \ . \label{am_bal}
\tag{S2}
\end{equation*}
Owing to the approximative nature of the expressions for  $\langle F_{\alpha} \rangle_\phi$ and   $ \langle F_{\lambda} \rangle_\phi$, we have replaced the mutual
friction parameter $\alpha (T)$ in Eq.~(\ref{am_bal}) with a new effective mutual-friction parameter $\alpha_\mathrm{am} $ for angular momentum, in which
$C_{\rm{am}}$ is a constant of order unity and the constant $\taam$ accounts for possible temperature-independent residual effects related to the angular-momentum
transfer. Solving Eq. (\ref{am_bal}) for $\Oms$ yields
\begin{equation*}
\Oms= \alpha_{\rm{am}} \Omega^2 \big / (  \alpha_{\rm{am}}\Omega+\lambda R^{-2})\ . \label{os}
\end{equation*}
In the approximation $\alpha\sb{am} \Rightarrow \alpha$,  this result reduces to the interpolation formula between the high and low temperature limits presented
in Ref.~[10]. Combining \emph{energy} and \emph{momentum balance} in Eqs.~(\ref{en}) and (S2) we get the model
\begin{equation*}  \frac{\vf}{\Omega R}=\frac{(\Cen\alpha+\taen)(C_{\rm{am}}\alpha+\taam)}{ C_{\rm{am}} \alpha+\taam   +\lambda\big  / ( \Omega R^2)} \; ,
\tag{S3} \label{EMB}
\end{equation*}
to which the measured data in Fig.~2 have been fitted.

\textbf{Bottleneck in energy accumulation of turbulent cascade.} The predictions of our model in Eq.~(\ref{EMB}), shown with solid lines in Fig.~2, are obtained
under the simplifying assumption that the fitting parameters are both temperature and rotation velocity independent. As seen in Fig.~2, the data show small
deviations from the solid lines (by about $\pm 5\%$) at intermediate temperatures: plateaus in the $T$-dependence of $\vf/(\Omega R)$ in the right panel, which
are most pronounced for $\Omega=0.5$\, rad/s at $T\simeq (0.32-0.35)T\sb c$. In Ref.~\cite{boue} it is shown that in this temperature range a bottleneck appears
in the energy accumulation rate at the scale of the inter-vortex distance, while at higher temperatures the Kelvin-wave cascade is still fully suppressed owing to
mutual friction. This shows up as a temperature dependence of $\taen$ which, however, is ignored in our model Eq.~(\ref{EMB}). In Ref.~[4] the bottleneck effect
was overestimated, as appropriate in a model with sharp crossover \cite{lvov1}. A more realistic model of gradual eddy-wave crossover \cite{lvov2}, which accounts
for the temperature-dependent suppression of the Kelvin-wave turbulence \cite{boue}, predicts a more modest dependence of $\taen$ on temperature, of the order of
a few percent. With these revisions, the small deviations of the data in Fig.~2 from the model in Eq.~(\ref{EMB}) might be attributed to the bottleneck effect.

\textbf{Numerical calculation.} Simulation calculations on vortex dynamics in the zero temperature limit present a numerical challenge, when performed with the
filament model, Biot-Savart integration along all vortices, and proper account of boundary conditions \cite{hanninen1}. At low mutual friction dissipation
Kelvin-wave excitations propagate to ever smaller length scales. Ultimately the turbulent energy cascade extends over many orders of magnitude to the smallest
scales approaching the superfluid coherence length, where other dissipation mechanisms might become important [14,30]. Thus with decreasing temperature the
demands on spatial resolution rapidly grow, if one wants to capture the essential features of the dynamics. Otherwise dissipation is underestimated and dominated
by numerical noise: insufficient resolution blocks the Kelvin cascade and creates an artificial bottleneck near the resolution limit \cite{hanninen2}. Typically
we find that in today's calculations the resolution limit $\Delta r$ compared to the characteristic geometric length $R$ of the problem is of order $10^{-2}$ --
$10^{-3}$ \cite{tsubota2012,baggaley2012}. To respond to Kelvin waves with wave length comparable to $\Delta r$, the time step in the calculations needs to be
$\Delta t \lesssim (\Delta r)^2 /\kappa$.

Owing to the numerical difficulties, reliable calculations with finite mutual friction are few in the $T \rightarrow 0$ limit. If they exist, they concentrate on
the effective dissipation in the presence of a vortex tangle. Our calculations on vortex front propagation [10] down to $T \sim 0.20\, T_\mathrm{c}$  are in
qualitative agreement with the measurements: the front velocity is two orders of magnitude higher than what the  single vortex estimate $V_f=\alpha R\Omega$
suggests, showing that dissipation is considerable. In contrast, in angular momentum transfer a break down is observed, as $\Omega_s$ drops well below $\Omega$.
Numerical simulations on the trapping of smooth spherical particles on quantized vortices also show that the trapping and thus the momentum transfer from the
particle to the vortex becomes much less efficient in the zero-temperature limit \cite{kivotides}. On the other hand, numerical calculations of the interaction
between similar particles and quantized vortices in an oscillating flow show some evidence for the drag effects at $T = 0$ \cite{fujiama}. Unfortunately, reliable
quantitative numerical results at very low temperatures  are still being worked on.\\

\end{document}